\documentclass[sigconf,screen]{acmart}

\usepackage{enumitem}
\usepackage{pifont}
\usepackage{multirow,makecell}
\usepackage{color,xcolor}
\usepackage{listings,amsfonts}
\usepackage{caption}
\usepackage{subcaption}
\usepackage{threeparttable}
\usepackage{bbding}
\usepackage{booktabs} 
\usepackage{longtable}
\usepackage{flushend}
\usepackage{tcolorbox}
\usepackage{bussproofs}
\usepackage{xspace}

\AtEndPreamble{
	\usepackage{hyperref}

}

\newcommand{\toolname}{\textsc{YASA}\xspace}

\author{Yayi Wang}
\authornote{Both authors contributed equally to this research.}
\email{wangyayi.wyy@antgroup.com}
\orcid{0009-0000-8880-1061}
\affiliation{%
  \institution{Ant Group}
  \city{Hangzhou}
  \state{Zhejiang}
  \country{China}
}

\author{Shenao Wang}
\authornotemark[1]
\email{shenaowang@hust.edu.cn}
\orcid{0000-0003-3818-3343}
\affiliation{%
  \institution{Huazhong University of Science and Technology}
  \city{Wuhan}
  \state{Hubei}
  \country{China}
}

\author{Jian Zhao}
\email{jian\_zhao\_@hust.edu.cn}
\orcid{0009-0003-5716-1462}
\affiliation{%
  \institution{Huazhong University of Science and Technology}
  \city{Wuhan}
  \state{Hubei}
  \country{China}
}

\author{Shaosen Shi}
\email{shishaosen.sss@antgroup.com}
\orcid{0009-0002-8031-8811}
\affiliation{%
  \institution{Ant Group}
  \city{Hangzhou}
  \state{Zhejiang}
  \country{China}
}

\author{Ting Li}
\email{jiting.lt@antgroup.com}
\orcid{0009-0000-4166-2455}
\affiliation{%
  \institution{Ant Group}
  \city{Hangzhou}
  \state{Zhejiang}
  \country{China}
}

\author{Yan Cheng}
\email{xiaoge.cy@antgroup.com}
\orcid{0009-0006-1025-5965}
\affiliation{%
  \institution{Ant Group}
  \city{Hangzhou}
  \state{Zhejiang}
  \country{China}
}

\author{Lizhong Bian}
\email{lizhong.blz@antgroup.com}
\orcid{0009-0006-0563-0409}
\affiliation{%
  \institution{Ant Group}
  \city{Hangzhou}
  \state{Zhejiang}
  \country{China}
}

\author{Kan Yu}
\authornote{Kan Yu~(kan.yk@antgroup) and Haoyu Wang~(haoyuwang@hust.edu.cn) are the co-corresponding authors.}
\email{kan.yk@antgroup.com}
\orcid{0009-0001-2554-7681}
\affiliation{%
  \institution{Ant Group}
  \city{Hangzhou}
  \state{Zhejiang}
  \country{China}
}

\author{Yanjie Zhao}
\email{yanjie_zhao@hust.edu.cn}
\orcid{0000-0001-8793-5367}
\affiliation{%
  \institution{Huazhong University of Science and Technology}
  \city{Wuhan}
  \state{Hubei}
  \country{China}
}

\author{Haoyu Wang}
\authornotemark[2]
\email{haoyuwang@hust.edu.cn}
\orcid{0000-0003-1100-8633}
\affiliation{%
  \institution{Huazhong University of Science and Technology}
  \city{Wuhan}
  \state{Hubei}
  \country{China}
}

\renewcommand{\shortauthors}{Yayi Wang et al.}

\begin{document}

\title{\toolname: Scalable Multi-Language Taint Analysis on the Unified AST at Ant Group}



\setlength{\textfloatsep}{5pt}
\setlength{\intextsep}{5pt}

\setlength{\abovecaptionskip}{5pt}
\setlength{\belowcaptionskip}{5pt} 


\begin{abstract}
Modern enterprises increasingly adopt diverse technology stacks with various programming languages, posing significant challenges for static application security testing (SAST). Existing taint analysis tools are predominantly designed for single languages, requiring substantial engineering effort that scales with language diversity. 
While multi-language tools like CodeQL, Joern, and WALA attempt to address these challenges, they face limitations in intermediate representation design, analysis precision, and extensibility, which make them difficult to scale effectively for large-scale industrial applications at Ant Group.
To bridge this gap, we present YASA (Yet Another Static Analyzer), a unified multi-language static taint analysis framework designed for industrial-scale deployment. 
Specifically, YASA introduces the Unified Abstract Syntax Tree (UAST) that provides a unified abstraction for compatibility across diverse programming languages.
Building on the UAST, YASA performs point-to analysis and taint propagation, leveraging a unified semantic model to manage language-agnostic constructs, while incorporating language-specific semantic models to handle other unique language features.
When compared to 6 single- and 2 multi-language static analyzers on an industry-standard benchmark, YASA consistently outperformed all baselines across Java, JavaScript, Python, and Go. In real-world deployment within Ant Group, YASA analyzed over 100 million lines of code across 7.3K internal applications. It identified 314 previously unknown taint paths, with 92 of them confirmed as 0-day vulnerabilities. All vulnerabilities were responsibly reported, with 76 already patched by internal development teams, demonstrating YASA's practical effectiveness for securing large-scale industrial software systems.
\end{abstract}



\maketitle

\section{Introduction}
Modern software systems increasingly adopt polyglot programming paradigms~\cite{lucas2024multi,mauer2017multi,philip2015multiple}, leveraging multiple programming languages across various layers and components to address rapidly evolving business and technological requirements~\cite{li2021understanding,li2021multi}. Recent surveys and empirical studies~\cite{tomassetti2014polyglotism,haoran24tse,wen23tosem} have revealed that over 80\% of large-scale applications utilize two or more programming languages in production environments.
As one of the world's leading financial technology platforms, Ant Group also relies on a diverse technology stack to support its business-critical applications. 
For example, microservices-based infrastructures often implement backend services in Java or Go due to their scalability and performance, while employing JavaScript, TypeScript, or popular frontend frameworks like React and Vue for user interface development~\cite{techstack}. 
This diversity in programming languages and frameworks enables organizations like Ant Group to harness the strengths of each technology for specific tasks, but it also introduces significant challenges in software development and maintenance~\cite{li2022multilingual,buro2020abstraction,sanaa2023multilanguage}.

Static Application Security Testing (SAST) has emerged as a fundamental approach for identifying vulnerabilities in software systems, with static taint analysis representing one of the most effective and widely adopted technique~\cite{arzt2014flowdroid,simon2009tajs,luo2022tchecker,shi2018pinpoint}. Extensive research from both academia~\cite{tan2023taie,shi2018pinpoint,li2022odgen,li2021proto} and industry~\cite{wang2020soa,xie2024codefuse,fabian2014joern} has demonstrated the efficacy of static analysis in detecting various taint-style vulnerabilities, including code/command injection~\cite{siddharth2023argus,yiu2023bimodal,ioannis2011phpaspis}, SQL injection~\cite{nenad2010taint}, deserialization~\cite{pang2025pfortifier,chen2024jdd}, server-side request forgery~(SSRF)~\cite{ji2025artemis,liu2025microservice}, and prototype pollution~\cite{li2021proto,li2022odgen,kang2023fast}. Despite these achievements, a critical challenge remains: are these static taint analysis tools truly engineered for industrial-scale, multi-language development practices?

\textbf{Research Gaps.}
One of the most critical limitations is that most academic taint analysis tools are designed as language-specific prototypes, lacking extensibility and portability. Representative tools such as FlowDroid~\cite{arzt2014flowdroid} and Tai-e~\cite{tan2023taie} for Java, SVF~\cite{sui2016svf,cheng2024csa} and Infer~\cite{infer} for C/C++, TAJS~\cite{simon2009tajs,laursen2024approximate}, DoubleX~\cite{fass2021doublex}, ODGen~\cite{li2022odgen,li2021proto,kang2023fast}, and GraphJS~\cite{ferreira2024graphjs} for JavaScript, PyT~\cite{pyt} and Pysa~\cite{pysa} for Python, Tchecker for PHP~\cite{luo2022tchecker,michael2017phpjoern}, and ARGOT~\cite{argot} for Go require dedicated engineering effort for each target language.
This approach of using language-specific analysis tools introduces several critical challenges that severely limit their applicability in industrial settings~\cite{checkmarx,seal2024sast}. 
First, development and maintenance costs are substantially increased because each language requires independent analysis frontends, dedicated dataflow engines, separate rule sets, and specialized expertise~\cite{christakis2016what,johnson2013why}. 
Secondly, the management and aggregation of analysis results become complicated. Outputs from different tools are often stored in incompatible formats~\cite{luo2019magpiebridge,johnson2013why,nachtigall2022criteria}, which prevents consolidation into a unified security knowledge base. 
Finally, as new programming languages and frameworks are adopted in production, building and maintaining separate analysis engines for each new stack is unsustainable~\cite{sadowski2018lessons}, resulting in coverage gaps and missed opportunities to secure the latest workflows.

In response, state-of-the-art industrial solutions have emerged to support multi-language taint analysis, most notably CodeQL~\cite{codeql,moor2007ql}, Joern~\cite{fabian2014joern,michael2017phpjoern}, WALA~\cite{wala,santos2022wala}, Infer~\cite{infer}, and Semgrep~\cite{semgrep}. However, despite their popularity and wide adoption, these frameworks face inherent limitations. First, from the architectural perspective, designing a unified intermediate representation (IR) that is both sufficiently expressive and extensible across multiple programming languages is intrinsically challenging~\cite{spanier2023ir,stanier2013ir}. 
This inherent difficulty in IR unification will be discussed further in \autoref{sec:ir}.
Second, these tools struggle to maintain high precision across languages, such as context-, path-, field-, and flow-sensitivity. Third, extensibility remains a challenge, as integrating new frameworks or language features into these multi-language platforms often demands substantial re-engineering effort~\cite{guo2024reactappscan}, limiting their adaptability in rapidly evolving software environments. As such, there remains an urgent research and engineering gap in designing scalable and precise taint analysis frameworks truly suited for the multi-language realities of modern enterprise software.

\textbf{Our Work.}
To address these challenges, we propose \textbf{YASA} (\textbf{\underline{Y}}et \textbf{\underline{A}}nother \textbf{\underline{S}}tatic \textbf{\underline{A}}nalyzer), a unified multi-language\footnote{In this paper, we focus on \textit{multi-language} analysis rather than \textit{cross-language} analysis.} static taint analysis framework. 
YASA is built on the Unified Abstract Syntax Tree (UAST), an intermediate representation that abstracts and normalizes core program constructs across languages. 
The key idea is that, despite substantial syntactic and semantic differences among programming languages, the fundamental mechanisms of taint propagation, including assignments, function calls, control flow, and data dependencies, can be modeled in a unified way while preserving language-specific semantics.
Based on UAST, YASA performs points-to analysis and taint propagation by combining reusable language-agnostic semantic models with language-specific semantic handlers for distinctive features, such as Python's duck typing and JavaScript's prototype chains.

\begin{figure*}[t]
\centering
\includegraphics[width=0.8\textwidth]{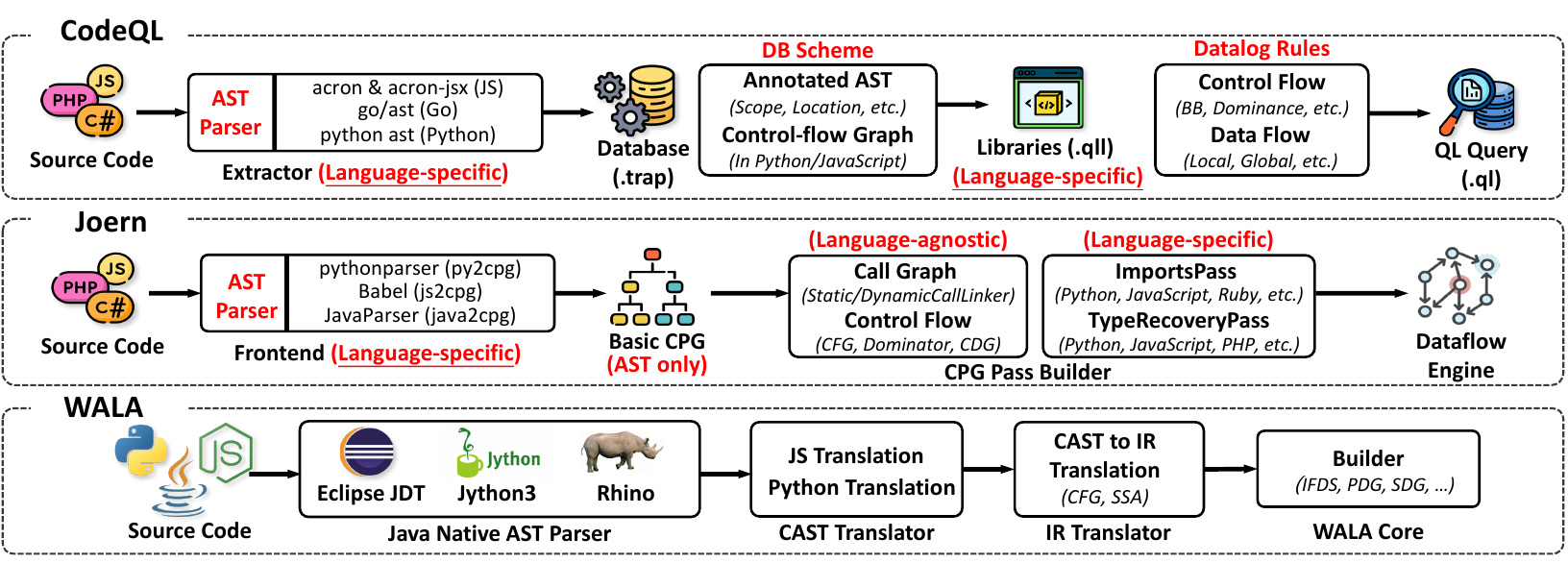}
\caption{Architectural Comparison of CodeQL, Joern, and WALA for Multi-language Taint Analysis.}
\label{fig:sota}
\end{figure*}

\textbf{Contributions.} To summarize, we make the following contributions in this paper:
\begin{itemize}[leftmargin=15pt]
\item \textbf{Unified Multi-Language IR Design.} We propose UAST, a novel representation to address the challenge of multi-language semantic unification. We have implemented frontend parsers for 4 programming languages, and both the specifications and these parsers have been fully open-sourced at \url{https://github.com/antgroup/YASA-UAST}.

\item \textbf{Scalable Taint Prototypes.} Building on UAST, we design and implement YASA, a multi-language taint analysis framework that achieves context-, path-, and field-sensitive taint propagation. This analyzer has been fully open-sourced at \url{https://github.com/antgroup/YASA-Engine}.

\item \textbf{Real-World Impact.} We analyze over 100 million lines of code across 7.3K applications at Ant Group. YASA has reported 314 previously unknown taint paths, with 92 of them confirmed as 0-day vulnerabilities. We have responsibly disclosed them, with 76 already patched by developers.
\end{itemize}

\section{Background and Motivation}
\subsection{Multi-language Taint Analysis}
\label{sec:multi}

Modern multi-language taint analysis frameworks such as CodeQL~\cite{codeql}, Joern~\cite{fabian2014joern}, and WALA~\cite{wala} adopt distinct approaches to support diverse programming languages. As illustrated in ~\autoref{fig:sota}, CodeQL employs a database-centric method, transforming source code into language-specific databases for analysis using declarative QL queries~\cite{avgustinov2016ql}. These databases capture annotated ASTs and semantic details, but require language-specific extractors and QL libraries (\texttt{.qll}). Joern, in contrast, uses a graph-based Code Property Graph (CPG)~\cite{fabian2014joern}, a unified intermediate representation combining AST, CFG, and PDG. Language-specific frontends generate CPGs, which are enriched with semantic passes for taint propagation. WALA relies on existing parsers and runtime environments (e.g., Rhino, Jython) to convert source code into a Common Abstract Syntax Tree (CAST), further processed into control flow graphs and SSA form for interprocedural analysis~\cite{santos2022wala}.

\begin{table}[t]
\centering
\fontsize{9}{12}\selectfont
\caption{\#LoC Distribution in CodeQL Between Language-Specific and Language-Agnostic Components}
\label{tab:codeql-stats}
\resizebox{0.9\linewidth}{!}{
\begin{tabular}{ccrr}
\toprule
\textbf{Category} & \textbf{Language} & \textbf{Extractor} & \textbf{Library} \\
\midrule
\multirow{6}{*}{\textbf{Language-Specific}} 
& JavaScript & 40,467 & 106,371 \\
& Java/Kotlin & 10,966$^{\dagger}$ & 69,118 \\
& C/C++ & N/A$^{\dagger}$ & 134,628 \\
& C\# & 31,590 & 54,301 \\
& Python & 26,995 & 67,003 \\
& Go & 8,604 & 31,157 \\
\cmidrule{2-4}
& \textbf{Total} & \textbf{118,622} & \textbf{462,578} \\
\midrule
\multirow{1}{*}{\textbf{Language-Agnostic}} 
& / & \multicolumn{2}{c}{27,492 (4.5\%)} \\
\bottomrule
\end{tabular}}
\begin{tablenotes}
\footnotesize
\item ${\dagger}$ The extractors for Java and C/C++ are not open source; only the Kotlin extractor is publicly available.
\end{tablenotes}
\end{table}

\noindent \textbf{Limitations in Practice.} Despite their strengths, these frameworks face notable limitations. CodeQL's reliance on language-specific extractors and database schemas results in significant fragmentation: as shown in ~\autoref{tab:codeql-stats}, only 4.5\% of its codebase is language-agnostic, making it costly to extend support for new languages. Joern's over-abstraction in CPG inflates representation size and reduces precision for dynamic language features like JavaScript's prototype chains and Python's runtime modifications~\cite{li2022odgen}. WALA, originally designed for Java bytecode, struggles with modern dynamic languages due to its dependence on interpreters (e.g., Rhino, Jython) and SSA-based representations, which fail to capture advanced semantics. 
These limitations will be discussed in \autoref{sec:ir}.

\subsection{Intermediate Representation} 
\label{sec:ir}
The choice of IR determines the effectiveness, precision, and extensibility of multi-language static analysis frameworks. Typical forms include AST, CFG, SSA, PDG, and their various cross-combinations. 

\begin{table}[t]
\centering
\caption{IRs Used in Single- and Multi-language SASTs.}
\label{tab:ir-table}
\resizebox{\linewidth}{!}{
\begin{tabular}{llcc}
\toprule
\textbf{Type} & \textbf{Tool: IR}        & \textbf{Vocabulary} & \textbf{Syntax} \\
\midrule
\multirow{10}{*}{\textbf{Single}}
    & SVF~\cite{sui2016svf}: SVF IR (LLVM IR)              & Register     & PAG, SSA, SVFG \\       
    & Tai-e~\cite{tan2023taie}: Tai-e IR (Jimple)          & Register     & SSA            \\
    & Doop~\cite{grech2017ptaint}: WALA/Soot IR            & Register     & SSA            \\
    & TAJS~\cite{simon2009tajs}: AST (jscomp, Babel)       & AST Node     & AST            \\
    & ODGen~\cite{li2022odgen}: AST (esprima)             & AST Node     & AST            \\
    & DoubleX~\cite{fass2021doublex}: AST (esprima)       & AST Node     & AST, CFG, PDG  \\
    & GraphJS~\cite{ferreira2024graphjs}: AST (estree)    & AST Node     & AST, CFG, MDG  \\
    & PyT~\cite{pyt}: AST~(python ast)                    & AST Node     & AST, CFG       \\
    & Pysa~\cite{pysa}: AST~(python ast)                  & AST Node     & AST, CFG       \\
    & ar-go-tools~\cite{argot}: Go SSA                    & Register     & SSA            \\
\midrule
\multirow{6}{*}{\textbf{Multi}}
    & CodeQL~\cite{codeql}: Database (.trap)              & DB Scheme    & AST, CFG       \\
    & CodeFuse-Query~\cite{xie2024codefuse}: Database           & DB Scheme    & AST, CFG       \\
    & Joern~\cite{fabian2014joern}: CPG                   & AST Node     & AST, CFG, PDG  \\
    & WALA~\cite{wala}: WALA IR                           & CAST \& Register & AST, CFG, SSA  \\
    & LLVM~\cite{llvm}: LLVM IR                           & Register     & SSA            \\
    & Infer~\cite{infer}: SIL                             & AST Node     & AST, CFG       \\
\bottomrule
\end{tabular}
}
\end{table}

\begin{figure*}[t]
\centering
\includegraphics[width=0.99\textwidth]{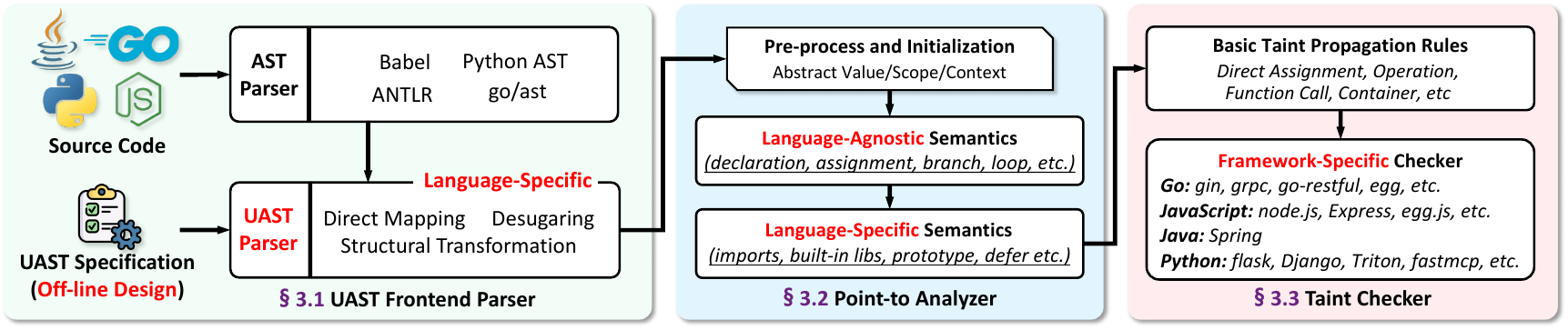}
\caption{Architecture Overview of YASA.}
\label{fig:yasa-overview}
\end{figure*}

\noindent\textbf{Single-Language IR Characteristics.} Single-language static analysis tools often employ IRs tailored to the semantic properties of their target languages, enabling high-precision analysis but sacrificing cross-language compatibility. As illustrated in ~\autoref{tab:ir-table}, compiled languages like C/C++ and Java predominantly favor register-based SSA forms that naturally align with their static compilation models. SVF leverages LLVM IR with specialized Pointer Assignment Graph (PAG) and Sparse Value-Flow Graph (SVFG) extensions to handle complex pointer relationships and memory operations characteristic of C/C++ programs~\cite{sui2016svf}. Similarly, Java-focused tools like Tai-e and Doop utilize register-based SSA representations (Tai-e IR and Jimple respectively) ~\cite{tan2023taie,grech2017ptaint}.
In contrast, dynamic languages such as JavaScript and Python favor AST-based representations that preserve high-level semantics critical for handling dynamic features like prototype chains, runtime attribute modification, and complex scoping rules. Tools like ODGen~\cite{li2022odgen}, DoubleX~\cite{fass2021doublex}, PyT~\cite{pyt}, and Pysa~\cite{pysa} rely on ASTs augmented with control flow information for effective analysis.

\noindent\textbf{Multi-Language IRs and Trade-offs.} Multi-language analysis frameworks face the challenge of balancing expressiveness, precision, and cross-language compatibility. As shown in ~\autoref{tab:ir-table}, existing approaches can be categorized into three primary strategies.

\textit{\underline{Database-relational unification}}, such as CodeQL~\cite{codeql} and CodeFuse-Query~\cite{xie2024codefuse}, transforms source code into queryable database schemas, which enables powerful declarative queries through relational database operations. However, in ~\autoref{sec:multi}, we demonstrate that each language requires independent schema design and analysis logic, leading to substantial engineering overhead that scales linearly with the number of supported languages.

\textit{\underline{Graph-based unification}}, represented by Joern's CPG, attempts to create a unified graph structure combining AST, CFG, and PDG information across languages~\cite{fabian2014joern}. While this approach provides intuitive graph traversal capabilities for analysis queries and enables some degree of cross-language analysis reuse, the CPG specification suffers from limitations due to its over-abstraction of high-level language features. The core issue lies in CPG's ``lowering'' strategy, where complex language constructs are decomposed into simpler, more primitive node combinations to fit the limited CPG vocabulary, leading to significant semantic information loss and representation bloat.
For example, a simple Python class definition node is decomposed in CPG into tens of nodes involving \texttt{TYPE\_DECL} nodes, synthetic \texttt{METHOD} nodes, and \texttt{<metaClassCallHandler>} for handling class semantics, and so on. 
Moreover, due to the limited vocabulary, the CPG specification fails to express some specific structures, such as \texttt{yield} expression in Python.

\textit{\underline{Low-level register-based unification}}, including WALA and LLVM, achieves cross-language compatibility by operating at reduced abstraction levels using register-based SSA representations~\cite{wala,llvm}. WALA transforms source code through language-specific interpreters into a common SSA form, enabling shared analysis algorithms across Java, JavaScript, and Python. LLVM IR operates at an even lower level, providing a register-based SSA representation that supports C/C++ and Go through their respective compiler frontends. While these approaches enable significant code reuse in analysis algorithms, they struggle with dynamic language features, where JavaScript's prototype, Python's runtime attribute modification, and dynamic method dispatch cannot be accurately captured in static SSA designed for traditional compiled languages.

\noindent\textbf{Insights \& Challenges.} Despite these limitations, existing multi-language approaches offer valuable insights into effective IR design. Notably, almost all successful multi-language frameworks incorporate AST as the foundational component: CodeQL extracts AST information into database schemas, Joern's CPG builds upon AST structures, and WALA begins with CAST before lowering to SSA. These practices suggest that AST-based representations offer advantages for multi-language unification by naturally preserving high-level semantic information. However, the critical challenge lies in determining the appropriate level of abstraction for cross-language compatibility. 
A key lesson learned from CPG is that over-abstraction with a limited vocabulary can unnecessarily complicate simple language constructs while failing to express certain language-specific features.
Therefore, successful AST-based multi-language IR requires achieving the balance between cross-language compatibility and language-specific semantic preservation. 
\section{Methodology}
We present YASA (Yet Another Static Analyzer), a scalable and extensible multi-language static taint analysis framework. Specifically, YASA utilizes the UAST, an offline-designed specification that unifies diverse language constructs into a common representation while preserving semantic fidelity. As shown in ~\autoref{fig:yasa-overview}, YASA consists of three key components. The \textit{UAST Frontend Parser} transforms language-specific ASTs into the UAST format through direct mapping, structural transformations, and desugaring. The \textit{Point-to Analyzer} performs pointer analysis by integrating language-agnostic semantics for common constructs such as assignments, branches, and loops, with language-specific extensions for unique features like Python decorators, Go's `defer`, and JavaScript's prototype chain. Finally, the \textit{Taint Checker} tracks taint flows using basic propagation rules (e.g., assignments, operations, function calls) and adapts to real-world frameworks such as Flask, Spring, and Express through the customized \textit{Framework-Specific Checker}. 

\subsection{UAST Frontend Parser}

\subsubsection{Design of UAST Specification.}
The core of YASA's multi-language unification lies in its UAST, which is designed to achieve both broad compatibility and precise semantic representation across diverse programming languages. To accomplish this, we conducted an in-depth analysis of the syntax and semantic constructs in multiple languages, identifying common patterns as well as language-specific peculiarities, which provided the foundation for constructing a unified node taxonomy. As shown in \autoref{tab:uast-nodes}, we classified AST nodes from different programming languages into three categories to heuristically design this specification: \textit{Universal Semantic Nodes}, \textit{Language-Specific Nodes}, and \textit{Reducible Nodes}.

\begin{table}[t]
\centering
\caption{Classification of AST Nodes and Language Constructs Across Different Programming Languages.}
\label{tab:uast-nodes}
\resizebox{\linewidth}{!}{%
\begin{tabular}{lll}
\toprule
\textbf{Category} & \textbf{AST Node} & \textbf{Description} \\
\midrule
\multirow{5}{*}{\textbf{Universal}} 
    & \texttt{Literal} & Constant values (e.g., numbers, strings) \\
    & \texttt{BinaryExpression} & Binary operations (e.g., +, -, ==) \\
    & \texttt{IfStatement} & Conditional branching \\
    & \texttt{CallExpression} & Function or method invocation \\
    & \texttt{RangeStatement} & Unified iteration construct \\
\midrule
\multirow{3}{*}{\textbf{Specific}}  
    & \texttt{YieldExpression} & Generator functions in Python \\
    & \texttt{ChanType} & Channel constructs in Go \\
    & \texttt{TupleExpression} & Tuple constructs in Python \\
\midrule
\multirow{4}{*}{\textbf{Reducible}}  
    & \texttt{[x for x in list]} & → \texttt{VariableDeclaration} + \texttt{RangeStmt} \\
    & \texttt{() => expr} & → \texttt{FunctionDefinition} \\
    & \texttt{Expression} & Root node for expressions in Python \\
    & \texttt{Comment} & Comment nodes in Go \\
\bottomrule
\end{tabular}%
}
\end{table}

\textit{Type\#1: Universal Semantic Nodes.}  
Universal semantic nodes represent constructs that share equivalent semantics across at least two supported languages, even if their syntax differs. These nodes enable consistent representation across languages and facilitate the reuse of analysis logic built on top of this unified structure. Examples include core programming constructs such as basic operations, control flow statements, and universal declarations. 
For instance, the \texttt{RangeStatement} unifies iteration constructs like JavaScript's \texttt{for-of} loops, Python's \texttt{for-in} syntax, and Go's \texttt{range} keyword into a single node type with shared semantics.

\textit{Type\#2: Language-Specific Nodes.}  
Language-specific nodes represent constructs that are unique to a particular language and cannot be easily expressed using universal semantic nodes without significant semantic loss. These nodes are explicitly retained in the UAST Specification to ensure accurate representation of language features that are critical for program analysis. 
Examples include Python's \texttt{YieldExpression} for generator functions, Go's \texttt{ChanType} for channel constructs, and Python's \texttt{TupleExpression} for tuple representations. 

\textit{Type\#3: Reducible Nodes.}  
Reducible nodes represent constructs that are either irrelevant to program analysis semantics or can be systematically reduced to equivalent combinations of other nodes without loss of meaning. These nodes are not retained in the UAST specification. 
For example, Python's list comprehensions (\texttt{[x for x in list]}) can be reduced to a sequence of \texttt{VariableDeclaration} and \texttt{RangeStatement} nodes, while JavaScript's arrow function expression (\texttt{() => expr}) can be transformed into \texttt{FunctionDefinition} nodes. Additionally, redundant nodes such as Python's \texttt{Expression} root node, which serves only as a wrapper, and Go's \texttt{Comment} node, will be discarded during the transformation process. 

\noindent\textbf{Overview of the UAST Specification.}  
As shown in~\autoref{tab:uast-spec}, the UAST specification encompasses 54 node types distributed across five primary syntactic categories: 4 basic nodes (e.g., literals, identifiers), 16 statement nodes (e.g., control flow, exception handling), 20 expression nodes (e.g., arithmetic operations, function calls), 4 declaration nodes (e.g., functions, classes, variables), and 10 type nodes (e.g., static and dynamic type systems). 

\begin{table}[t]
\centering
\caption{UAST Node Specification by Syntactic Category.}
\label{tab:uast-spec}
\resizebox{\linewidth}{!}{%
\begin{tabular}{llrr}
\toprule
\textbf{Category} & \textbf{Node Examples} & \textbf{\#UNI} & \textbf{\#SPEC} \\
\midrule
\multirow{3}{*}{\textbf{Basic Nodes (4)}}  
    & \texttt{Noop} & \multirow{3}{*}{4} & \multirow{3}{*}{0} \\
    & \texttt{Literal} &  &  \\
    & \texttt{Identifier} &  &  \\
\midrule
\multirow{4}{*}{\textbf{Statement Nodes (16)}}  
    & \texttt{IfStatement} & \multirow{5}{*}{15} & \multirow{5}{*}{1} \\
    & \texttt{ReturnStatement} &  &  \\
    & \texttt{WhileStatement} &  &  \\
    & \texttt{RangeStatement} &  &  \\
\midrule
\multirow{5}{*}{\textbf{Expression Nodes (20)}}  
    & \texttt{BinaryExpression} & \multirow{5}{*}{11} & \multirow{5}{*}{9} \\
    & \texttt{CallExpression} &  &  \\
    & \texttt{MemberAccess} &  &  \\
    & \texttt{NewExpression} &  &  \\
    & \texttt{YieldExpression} &  &  \\
\midrule
\multirow{4}{*}{\textbf{Declaration Nodes (4)}}  
    & \texttt{FunctionDefinition} & \multirow{4}{*}{3} & \multirow{4}{*}{1} \\
    & \texttt{VariableDeclaration} &  &  \\
    & \texttt{ClassDefinition} &  &  \\
    & \texttt{PackageDeclaration} &  &  \\
\midrule
\multirow{4}{*}{\textbf{Type Nodes (10)}}  
    & \texttt{PrimitiveType} & \multirow{5}{*}{2} & \multirow{5}{*}{8} \\
    & \texttt{ArrayType} &  &  \\
    & \texttt{PointerType} &  &  \\
    & \texttt{ChanType} &  &  \\
\midrule
\textbf{Total~(54)} & / & 35 & 19 \\
\bottomrule
\end{tabular}%
}
\end{table}

\subsubsection{AST to UAST Transformation.}
To transform language-specific ASTs into the UAST representation, we developed three distinct transformation rules.

\textit{Rule\#1: Direct Mapping Rules.}  
Direct mapping rules handle AST nodes that correspond directly to preserved language-specific nodes or universal nodes. For language-specific nodes, UAST retains their syntax and semantics. Examples include Python's \texttt{ast.Yield} → \texttt{YieldExpression}, which is retained as it represents unique constructs. 
For universal nodes, direct mapping allows for straightforward conversions without structural modifications, such as Python's \texttt{ast.If} → \texttt{IfStatement}.

\textit{Rule\#2: Structural Transformation Rules.}  
Structural transformation rules address AST nodes that map to universal nodes in the UAST but require aggregation or restructuring to unify cross-language variations. These rules transform language-specific structural patterns into semantically equivalent UAST representations. Examples include \texttt{RangeStatement}, which serves as a universal representation for loop constructs across languages, such as Python's \texttt{ast.For}, JavaScript's for-of, Java's enhanced for, and Go's range constructs. 

\textit{Rule\#3: Desugaring Rules.}  
Desugaring rules systematically decompose language-specific syntactic sugar that corresponds to reducible nodes, transforming high-level constructs into equivalent combinations of invariant, aggregated, or preserved UAST nodes. These rules normalize syntactic convenience features into their underlying computational semantics, simplifying representation while preserving semantics. Examples include Python's list comprehensions (\texttt{ast.ListComp}) desugared into \texttt{Sequence} nodes, represented as a combination of \texttt{VariableDeclaration} for temporary variables, \texttt{RangeStatement} for iteration, and a final reference to the temporary variable. 

\subsection{Point-to Analyzer}
The \textit{Point-to Analyzer} is the core computational engine of YASA, implementing context-sensitive, path-sensitive, and field-sensitive point-to analysis over the unified UAST. The analyzer is designed to handle both language-agnostic semantics and language-specific features through a unified abstraction framework.

\subsubsection{Problem Formulation}
To enable precise interprocedural point-to analysis across multiple languages, YASA defines three key abstractions: abstract value domain, variable scope, and context state.

\textbf{Abstract Value Domain.} 
The core abstraction in YASA is the value domain \(\mathcal{V}\), which models the abstract values that variables may reference across different programming languages:
$$
\mathcal{V} ::= \mathcal{P}rim(\tau, v) \mid \mathcal{O}bj(\mathcal{H}) \mid \mathcal{S}ym(\tau, n) \mid \mathcal{P}hi(\mathcal{T})
$$
Here, \(\mathcal{P}rim(\tau, v)\) denotes primitive values with type \(\tau\) and concrete value \(v\); \(\mathcal{O}bj(\mathcal{H})\) denotes objects and structured values, where \(\mathcal{H}: \mathcal{F}ield \rightharpoonup \mathcal{V}\) is a field map for field-sensitive analysis; \(\mathcal{S}ym(\tau, n)\) denotes symbolic values whose concrete references cannot be statically resolved; and \(\mathcal{P}hi(\mathcal{T})\) denotes path-sensitive values represented by a tree \(\mathcal{T} = (N, E, \lambda)\), where nodes store values, edges encode control-flow alternatives, and \(\lambda: E \rightarrow \mathcal{P}ath\) annotates edges with branch conditions.

\textbf{Variable Scope Abstraction.} 
YASA models program scopes to capture lexical scoping relationships and variable state:
$$
\mathcal{S}cope = \langle \rho, \delta, \sigma_p, \tau_s \rangle
$$
where \(\rho: \mathcal{V}ar \rightharpoonup \mathcal{V}\) maps variables to abstract values, \(\delta: \mathcal{V}ar \rightharpoonup \mathcal{D}eclInfo\) stores declaration metadata, \(\sigma_p \in \mathcal{S}cope \cup \{\bot\}\) points to the parent scope, and \(\tau_s \in \{\mathcal{S}cope, \mathcal{F}clos, \mathcal{G}lobal, \mathcal{U}ninit, \mathcal{S}ym\}\) identifies the scope kind. 
These scope kinds represent, respectively, ordinary lexical scopes, function closures with captured environments, global scopes, uninitialized bindings, and symbolic scopes for unresolved references.

\textbf{Context State Abstraction.} 
To support interprocedural and path-sensitive analysis, YASA maintains the execution context:
$$
\mathcal{C}tx = \langle \sigma, \kappa, \pi \rangle
$$
where \(\sigma \in \mathcal{S}cope\) is the current scope, \(\kappa \in \mathcal{S}cope^*\) is the call stack represented as a sequence of function closures, and \(\pi \in \mathcal{P}ath^*\) records the branch conditions along the current execution path.

\subsubsection{Language-Agnostic Operational Semantics}
YASA implements context-sensitive, field-sensitive, and path-sensitive point-to analysis over the unified domain. We define the operational semantics using Structural Operational Semantics (SOS) rules that maintain precision across different sensitivity dimensions while enabling cross-language analysis reuse.

{\vspace{-2.5ex}
\small
\begin{gather*}
\underset{\text{(Context Sensitivity)}}{
  \frac{
    \langle f(e_1, \ldots, e_n), \mathcal{C}tx \rangle \Downarrow \langle \mathcal{F}clos, [v_1, \ldots, v_n] \rangle \quad \kappa' = \kappa \cdot [\mathcal{F}clos] \land |\kappa'| \leq k
  }{
    \langle x :^i f(e_1, \ldots, e_n), \mathcal{C}tx \rangle \Downarrow \langle \rho[x \mapsto v_{\text{ret}}], \mathcal{C}tx' \rangle
  }
}
\\[1ex]
\underset{\text{(Path Sensitivity)}}{
  \frac{
    \langle e, \mathcal{C}tx \rangle \Downarrow \langle v_{\text{test}}, \mathcal{C}tx' \rangle \quad \text{evaluate}(v_{\text{test}}, \pi) = \text{Unknown}
  }{
    \langle \text{if}(e)\{s_1\}\text{else}\{s_2\}, \mathcal{C}tx \rangle \Downarrow \langle \text{mergeContexts}(\mathcal{C}tx_1, \mathcal{C}tx_2) \rangle
  }
}
\\[1ex]
\underset{\text{(Field Sensitivity)}}{
  \frac{
    \langle e, \mathcal{C}tx \rangle \Downarrow \langle \mathcal{O}bj(\mathcal{H}), \mathcal{C}tx' \rangle \quad v = \mathcal{H}(f)
  }{
    \langle x :^i e.f, \mathcal{C}tx \rangle \Downarrow \langle \rho[x \mapsto v], \mathcal{C}tx' \rangle
  }
}
\end{gather*}}

\textbf{Context Sensitivity.} YASA maintains calling context through state cloning and call stack extension, enabling precise distinction of function behavior under different calling contexts.
The rule maintains calling context through call stack extension $\kappa' = \kappa \cdot [\mathcal{F}clos]$ where each function call creates a new execution state with extended context. Bounded call stack depth prevents infinite recursion while preserving analysis precision.

\textbf{Path Sensitivity.} YASA tracks execution paths by extending path conditions and merging results from different branches, enabling precise reasoning about program behavior under different execution scenarios.
The rule implements path-sensitive analysis by forking execution states when branch conditions cannot be statically determined. Each branch extends path conditions $\pi_T = \pi :: v_{\text{test}}$ and $\pi_F = \pi :: \neg v_{\text{test}}$ to maintain constraints along different execution paths. The $\text{mergeContexts}$ operation combines results using $\mathcal{P}hi(\mathcal{T})$ values that preserve path-specific information through tree structures.

\textbf{Field Sensitivity.} YASA maintains field-level precision by tracking individual object properties separately, avoiding the imprecision that results from field-insensitive approaches where all fields are merged.
The rule maintains field-level precision through the field mapping $\mathcal{H}: \mathcal{F}ield \rightharpoonup \mathcal{V}$ within object values $\mathcal{O}bj(\mathcal{H})$. Field access operations retrieve values through precise field lookups $\mathcal{H}(f)$, while field assignments create updated object values with modified mappings $\mathcal{H}[f \mapsto v]$. 

\subsubsection{Language-Specific Operational Semantics}
While the \textit{Unified Semantic Analyzer} handles most program constructs through universal UAST nodes, some language-specific features require specialized treatment to preserve analysis precision. YASA addresses these cases through \textit{Language-Specific Semantic Handlers}, which extend the unified analysis with targeted rules for distinctive language semantics.

{\small
\begin{gather*}
  \underset{\text{(Python)}}{
    \frac{
      \langle \text{class } C(\text{Base}), \mathcal{C}tx \rangle \Downarrow \langle \text{scope}(C), \mathcal{C}tx' \rangle \quad \text{Base}.\text{field}[f] = v
    }{
      \langle C.\text{inherit}(\text{Base}), \mathcal{C}tx \rangle \Downarrow \langle C.\text{field}[f \mapsto v'], \mathcal{C}tx' \rangle
    }
  }
  \\[2ex]
  \underset{\text{(JavaScript)}}{
    \frac{
      \langle \text{obj.method}(), \mathcal{C}tx \rangle \quad \text{obj}.\text{\_\_proto\_\_}.\text{method} \neq \emptyset
    }{
      \langle \text{resolve}(\text{method}), \mathcal{C}tx' \rangle \Downarrow \langle \text{method}.\text{call}(\text{obj}), \mathcal{C}tx' \rangle
    }
  }
  \\[2ex]
  \underset{\text{(Java)}}{
    \frac{
      \langle \text{@Data class } C, \mathcal{C}tx \rangle \Downarrow \langle \text{scope}(C), \mathcal{C}tx' \rangle \quad C.\text{field}[f] \neq \emptyset
    }{
      \langle \text{@Data}, \mathcal{C}tx' \rangle \Downarrow \langle C[\text{get}F(), \text{set}F()], \mathcal{C}tx'' \rangle
    }
  }
  \\[2ex]
  \underset{\text{(Go)}}{
    \frac{
      \langle \text{interface } I\{m()\}, \mathcal{C}tx \rangle \Downarrow \langle \text{interface}(I), \mathcal{C}tx' \rangle \quad T.\text{methods}[m] \neq \emptyset
    }{
      \langle T \text{ implements } I, \mathcal{C}tx' \rangle \Downarrow \langle T \subseteq I, \mathcal{C}tx'' \rangle
    }
  }
\end{gather*}
}

Due to space limitations and the diversity of language features, the rules above illustrate one representative feature for each language. The \textit{Python handler} models inheritance and field propagation across class hierarchies. The \textit{JavaScript handler} resolves prototype-based method dispatch. The \textit{Java handler} captures annotation-driven code generation, such as Lombok-generated accessors. The \textit{Go handler} models structural interface satisfaction based on method sets. Together, these handlers complement the unified semantics by covering language behaviors that cannot be fully represented through universal rules alone.

\subsection{Taint Checker}
The \textit{Taint Checker} in YASA is designed as a highly extensible, plugin-based system for detecting vulnerabilities through taint propagation. Specifically, checkers are implemented as independent, event-driven plugins, allowing users to customize and extend the analysis without modifying core logic. These checkers include \textit{Basic Taint Propagation Rules} and the \textit{Framework-Specific Checker}.

\subsubsection{Basic Taint Propagation Rules}
The taint analysis in YASA is built upon a modular set of taint propagation rules that define how taint flows between program elements during execution. As summarized in \autoref{fig:taint-propagation-rules}, we provide the core propagation rules in YASA. These include basic assignment propagation, container field interactions, function call propagation, and language-specific extensions for JavaScript and Go. For instance, JavaScript-specific rules capture taint propagation through prototypes and promises, while Go-specific rules handle taint flow via channels. The modular design of YASA allows developers to extend or customize these rules to address specific analysis requirements, ensuring flexibility in supporting diverse programming paradigms. 

\begin{figure}[t]
\centering
\fbox{
\begin{minipage}{0.95\linewidth}
\footnotesize 
\textbf{Notations:}
\begin{itemize}[leftmargin=15pt]
    \item $v, o \in \mathcal{V}$: Abstract values or objects.
    \item $f \in \mathcal{F}$: Field names in objects.
    \item $o_2 \prec o_1$: Object $o_2$ inherits from object $o_1$.
    \item $\tau_t(v) \in \{\top, \bot\}$: Taint state of $v$, $\top$ means tainted, and $\bot$ means not.
    \item $s$: A program statement (e.g., assignments, function calls, etc.).
    \item $\rightarrow$: Denotes taint propagation.
\end{itemize}

\textbf{Rules:}
\setlength{\abovedisplayskip}{2.5pt} 
\setlength{\belowdisplayskip}{2.5pt} 
\begin{enumerate}[leftmargin=15pt, label=\arabic*.]
    \item \textbf{Assignment Propagation:}
    \[
    s : x = y, \; \tau_t(y) = \top 
    \quad \rightarrow \quad 
    \tau_t(x) = \top
    \]

    \item \textbf{Container Field Propagation:}
    \[
    s : o.f = v, \; \tau_t(v) = \top 
    \quad \rightarrow \quad 
    \tau_t(o.f) = \top
    \]
    \[
    s : x = o.f, \; \tau_t(o.f) = \top 
    \quad \rightarrow \quad 
    \tau_t(x) = \top
    \]

    \item \textbf{Function Call Propagation:}
    \[
    s : z = f(x), \; \tau_t(x) = \top, \;  \tau_t(\text{return}(f)) = \top
    \quad \rightarrow \quad 
    \tau_t(z) = \top
    \]

    \item \textbf{Prototype Propagation~(JavaScript):}
    \[
    s : o_1.\text{prototype}.f = v, \; \tau_t(v) = \top 
    \quad \rightarrow \quad
    \tau_t(o_2.f) = \top, \; \forall o_2 \; (o_2 \prec o_1)
    \]
    \[
    s : x = o_1.\text{prototype}.f, \; \tau_t(o_1.\text{prototype}.f) = \top 
    \quad \rightarrow \quad 
    \tau_t(x) = \top
    \]

    \item \textbf{Promise Propagation~(JavaScript):}
    \[
    s : p = \text{Promise.resolve}(v), \; \tau_t(v) = \top 
    \quad \rightarrow \quad 
    \tau_t(p) = \top
    \]
    \[
    s : q = p.\text{then}(\text{callback}), \; \tau_t(p) = \top 
    \quad \rightarrow \quad 
    \tau_t(q) = \top
    \]
    \[
    s : v = \text{await } p, \; \tau_t(p) = \top 
    \quad \rightarrow \quad 
    \tau_t(v) = \top
    \]

    \item \textbf{Channel Propagation~(Go):}
    \[
    s : ch \leftarrow v, \; \tau_t(v) = \top 
    \quad \rightarrow \quad 
    \tau_t(ch) = \top
    \]
    \[
    s : v = \text{<-} ch, \; \tau_t(ch) = \top 
    \quad \rightarrow \quad 
    \tau_t(v) = \top
    \]
    \[
    s : ch_2 \leftarrow (\text{<-} ch_1), \; \tau_t(ch_1) = \top 
    \quad \rightarrow \quad 
    \tau_t(ch_2) = \top
    \]
\end{enumerate}
\end{minipage}
}
\caption{Taint Propagation Rules in YASA.}
\label{fig:taint-propagation-rules}
\end{figure}

\subsubsection{Framework-Specific Checker}

Framework-specific checkers in YASA are designed to handle taint propagation patterns unique to specific libraries, frameworks, or APIs. These checkers extend the basic taint propagation rules to account for framework-specific abstractions, such as routing mechanisms, data bindings, and middleware pipelines. For instance, in web frameworks, taint tracking often requires identifying how user inputs flow through routing logic and into application handlers. This includes detecting user-controllable input sources encapsulated by the framework and accurately simulating taint propagation through framework-defined structures.
YASA implements framework-specific checkers that focus on three main tasks: performing taint propagation within framework-specific execution flows, identifying entry points where taint can begin, and defining framework-specific sources to capture user-controlled inputs. For example, in frameworks like SpringMVC, the checker identifies controller methods as entry points and their parameters as sources. Similarly, in Egg.js, the \texttt{EggAnalyzer} models its unconventional module addressing and routing mechanisms, allowing the checker to track taint from sources such as \texttt{this.ctx.request}. 
By leveraging YASA’s modular design, these checkers can be extended or customized to support additional frameworks or APIs, ensuring that YASA remains adaptable to the evolving ecosystem of modern software development. 
\section{Evaluation}

To demonstrate the effectiveness, extensibility, and practicality, we conducted experiments addressing three research questions (RQs):

\textbf{RQ1: Detection Effectiveness.} How does YASA's detection performance compare to existing single-language and multi-language static analysis tools across different programming languages?

\textbf{RQ2: Language Extensibility.} How effectively does YASA's universal design reduce the engineering effort required for integrating each programming language?

\textbf{RQ3: Real-world Practicality.} What is the practical impact of YASA when deployed in production environments for analyzing large-scale multi-language codebases?

\subsection{Evaluation Setup}

\noindent \textbf{Implementation.} We have implemented a prototype of YASA with over 47,000 LoC, excluding any third-party libraries or open-source tools. So far, YASA supports 4 mainstream programming languages used at Ant Group: Python, JavaScript, Java, and Go, covering a total of 16 framework-specific checkers with dedicated taint analysis rules.
In Python, YASA supports frameworks such as Flask, Django, and FastAPI. For JavaScript, we provide checkers for frameworks like Express, Egg.js, and Node.js. In Java, YASA includes rules tailored for Spring. For Go, YASA supports popular frameworks like Gin, gRPC, Beego, and Gorilla Mux. 

\noindent \textbf{Running Environment}  
All experiments run on an Alibaba Cloud Elastic Compute Service (ECS) instance (\texttt{ecs.c6a.16xlarge}). The server operates on Alibaba Cloud Linux 3 and is equipped with 64 vCPUs, 128 GB of RAM, and 4 data disks, each with 8 TB capacity, providing a total storage of 32 TB.

\begin{table}[t]
\centering
\caption{Overview of xAST Benchmark Test Cases.}
\label{tab:xast-benchmark}
\begin{tabular}{lccc}
\toprule
\textbf{Language} & \textbf{\#Soundness${^\dagger}$} & \textbf{\#Completeness${^\ddagger}$} & \textbf{\#Total Cases} \\
\midrule
JavaScript & 134 & 49 & 183 \\
Python & 252 & 74 & 326 \\
Java & 111 & 58 & 169 \\
Go & 105 & 68 & 173 \\
\midrule
\textbf{Total} & \textbf{602} & \textbf{249} & \textbf{851} \\
\bottomrule
\end{tabular}
\begin{tablenotes}[leftmargin=1pt]
\footnotesize
\item ${^\dagger}$ \textit{\textbf{\#Soundness}}: Test cases evaluating support for language-specific features (e.g., prototype chains or async functions in JavaScript). \\
\item ${^\ddagger}$ \textit{{\textbf{\#Completeness}}}: Test cases assessing context-, field-, flow-, and path-sensitivity.
\end{tablenotes}
\end{table}

\begin{table*}[t]
\centering
\fontsize{7.5}{10}\selectfont
\caption{Detection Performance Comparison on xAST Benchmark.}
\label{tab:detection-performance}
\begin{tabular}{llcccccccc}
\toprule
\multirow{2}{*}{\textbf{Type}} & \multirow{2}{*}{\textbf{Tool}} & \multicolumn{2}{c}{\textbf{Java}} & \multicolumn{2}{c}{\textbf{Go}} & \multicolumn{2}{c}{\textbf{JavaScript}} & \multicolumn{2}{c}{\textbf{Python}} \\
\cmidrule(lr){3-4} \cmidrule(lr){5-6} \cmidrule(lr){7-8} \cmidrule(lr){9-10}
& & \textbf{Sound} & \textbf{Complete} & \textbf{Sound} & \textbf{Complete} & \textbf{Sound} & \textbf{Complete} & \textbf{Sound} & \textbf{Complete} \\
\midrule
\multicolumn{2}{c}{\textbf{Benchmark}} & 111 & 58 & 105 & 68 & 134 & 49 & 252 & 74 \\
\midrule
\multirow{6}{*}{\textbf{Single}} 
& Doop & 59 (53\%) & 25 (43\%) & - & - & - & - & - & - \\
& Tai-e & 75 (68\%) & 24 (41\%) & - & - & - & - & - & - \\
& ARGOT & - & - & 67 (64\%) & 38 (56\%) & - & - & - & - \\
& DoubleX & - & - & - & - & 32 (24\%) & 15 (31\%) & - & - \\
& ODGen & - & - & - & - & 60 (45\%) & 19 (39\%) & - & - \\
& PySA & - & - & - & - & - & - & 138 (55\%) & 34 (46\%) \\
\midrule
\multirow{2}{*}{\textbf{Multi}} 
& CodeQL & 55 (50\%) & 21 (36\%) & 63 (60\%) & 38 (56\%) & 88 (66\%) & 25 (51\%) & 120 (48\%) & 30 (41\%) \\
& Joern & 70 (63\%) & 17 (29\%) & 54 (51\%) & 13 (19\%) & 79 (59\%) & 18 (37\%) & 125 (50\%) & 25 (34\%) \\
\midrule
\textbf{Ours} & \textbf{YASA} & \textbf{80 (72\%)} & \textbf{32 (55\%)} & \textbf{96 (91\%)} & \textbf{45 (66\%)} & \textbf{118 (88\%)} & \textbf{35 (71\%)} & \textbf{176 (70\%)} & \textbf{44 (59\%)} \\
\multicolumn{2}{c}{\textbf{Improvement}} & \textcolor{red}{\textbf{4$\sim$22\%$\uparrow$}} & \textcolor{red}{\textbf{14$\sim$26\%$\uparrow$}} & \textcolor{red}{\textbf{27\%$\sim$40\%$\uparrow$}} & \textcolor{red}{\textbf{10$\sim$47\%$\uparrow$}} & \textcolor{red}{\textbf{22$\sim$64\%$\uparrow$}} & \textcolor{red}{\textbf{20\%$\sim$40\%$\uparrow$}} & \textcolor{red}{\textbf{15$\sim$22\%$\uparrow$}} & \textcolor{red}{\textbf{13$\sim$25\%$\uparrow$}} \\
\bottomrule
\end{tabular}
\end{table*}

\noindent \textbf{Benchmark.} 
To evaluate YASA, we considered several existing taint analysis benchmarks. Macro-benchmarks such as OWASP Benchmark~\cite{owaspbench} and SecBench.js~\cite{bhuiyan2023secbenchjs} provide realistic scenarios, but are often biased by predefined taint specifications and benchmark-specific vulnerability patterns. In contrast, micro-benchmarks such as TaintBench~\cite{luo2022taintbench} and DroidBench~\cite{steven2014droidbench} focus on Android, which YASA does not currently support\footnote{YASA currently does not support Android.}, while Juliet Test Suite~\cite{juliet13} and SecuriBench-Micro~\cite{securibenchmicro} are restricted to specific languages. 
We therefore use xAST\footnote{The ``x'' in xAST indicates support for multiple forms of application security testing, including SAST, IAST, and DAST.}~\cite{xastbench}, an industrial multi-language micro-benchmark for taint analysis. xAST includes soundness cases for language-specific features and completeness cases for sensitivity dimensions such as context-, field-, flow-, object-, and path-sensitivity. Each test case includes both positive and negative examples, reducing the risk of inflated performance from simple pattern matching. As shown in ~\autoref{tab:xast-benchmark}, xAST covers four languages, with 602 soundness cases and 249 completeness cases.

\noindent \textbf{Baseline.} To evaluate RQ1, we select representative single-language and multi-language static taint analysis tools as baselines. For single-language comparisons, we choose SOTA tools for each programming language supported by YASA. For Java, we compare against Tai-e~\cite{tan2023taie} and Doop~\cite{grech2017ptaint}, both of which are widely-recognized academic static analysis frameworks. For JavaScript, we evaluate against DoubleX~\cite{fass2021doublex} and ODGen~\cite{li2022odgen}. For Python, we compare with Pysa~\cite{pysa}, an industry-standard static analyzer from Facebook. For Go analysis, we compare against ARGOT~\cite{argot}, a collection of analysis tools for Go developed by AWSLabs. For multi-language baseline comparisons, we select CodeQL~\cite{codeql} and Joern~\cite{fabian2014joern}. We do not include Semgrep~\cite{semgrep} because the community version of Semgrep does not support inter-procedural taint analysis.

\subsection{RQ1: Detection Effectiveness}
To evaluate YASA's detection effectiveness, we compare its performance against six SOTA single-language tools and two multi-language frameworks on the xAST benchmark across four programming languages. ~\autoref{tab:detection-performance} presents comprehensive results for both soundness cases and completeness cases.

\noindent\textbf{Comparison with Single-Language Tools.} YASA demonstrates superior performance compared to all single-language tools in both soundness and completeness. For Java, YASA passes 52 soundness cases and 40 completeness cases, compared to Tai-e's 49 and 30, and Doop's 38 and 31, respectively. Specifically, both Tai-e and Doop fail in field-sensitive analysis for arrays and maps, treating partial contamination as whole-object pollution, and struggle with path-sensitive scenarios where they conservatively merge execution paths rather than distinguishing feasible branches. 
In Go analysis, YASA passes 96 soundness cases and 45 completeness cases, significantly outperforming ARGOT, which passes 68 and 38 cases, respectively. ARGOT's limitations include its conservative over-approximation, lack of automatic taint clearing on variable reassignment, and coarse control flow handling for defer statements and concurrency constructs.  
For JavaScript, YASA passes 118 soundness cases and 35 completeness cases, compared to ODGen's 60 and 19, and DoubleX's 32 and 15. DoubleX fails to support modern JavaScript features such as asynchronous programming, ES6+ syntax, and cross-module analysis. ODGen performs better but still struggles with advanced object-oriented features and modern constructs.  
In Python, YASA passes 176 soundness cases and 44 completeness cases, compared to PySA's 138 and 34. PySA exhibits gaps in analyzing Python-specific features, including list comprehensions, generator functions, exception handling, and dynamic typing (e.g., aliasing and reflection mechanisms).

\noindent\textbf{Comparison with Multi-Language Tools.} YASA consistently outperforms both CodeQL and Joern across all languages in soundness and completeness. For CodeQL, the number of passed soundness cases ranges from 51 in Python to 88 in JavaScript, while completeness cases range from 28 in Java to 37 in Go. These variations arise from CodeQL's language-specific extractor architecture, where independent implementations for each language lead to inconsistent support. CodeQL struggles with context-sensitive analysis for Java parameter passing, field-sensitive analysis for JavaScript data structures, and Python-specific constructs such as list comprehensions and dynamic typing. Additionally, it fails to handle Go's concurrency primitives, such as goroutines and channels, and nested struct analysis.  
Joern exhibits even greater inconsistencies, passing only 23 to 65 soundness cases and 11 to 22 completeness cases across languages. It struggles with polymorphism analysis, failing to effectively handle inheritance hierarchies, method overriding, and interface implementations. Joern also has significant gaps in field-sensitive analysis for multi-dimensional arrays, nested objects, and container types, often conflating partial contamination with whole-object pollution. The tool lacks support for modern language features, such as ES6+ constructs in JavaScript, Python generators and decorators, and Go's concurrency mechanisms. Furthermore, Joern fails to perform cross-module analysis, making it unable to track taint flows across file and module boundaries, significantly limiting its utility for multi-language projects.

\subsection{RQ2: Language Extensibility}
To evaluate YASA's language extensibility, we assess how effectively its unified design reduces engineering complexity when adding support for new programming languages. Our evaluation focuses on two key aspects: engineering efforts required to extend YASA to new languages, and the effectiveness of the language-agnostic semantic analyzer when applied without language-specific adjustments. 

\begin{table}[t]
\centering
\fontsize{8}{10}\selectfont
\setlength{\tabcolsep}{5pt} 
\caption{Semantic Rules Distribution Across Languages.}
\label{tab:semantic-reuse}
\begin{tabular}{lccc}
\toprule
\multirow{2}{*}{\textbf{Language}} & \multicolumn{2}{c}{\textbf{\#Semantic Rules}} & \multirow{2}{*}{\textbf{Reuse Rate}} \\
\cmidrule(lr){2-3} 
 & \textbf{Agnostic} & \textbf{Specific} & \\
\midrule
JavaScript & \multirow{4}{*}{52} & 19 & 73.2\% \\
Python     &                     & 15 & 77.6\% \\
Java       &                     & 10 & 83.9\% \\
Go         &                     & 18 & 74.3\% \\
\midrule
\textbf{Average} & \textbf{-} & \textbf{15.5} & \textbf{77.3\%} \\
\bottomrule
\end{tabular}%
\end{table}

\begin{table}[t]
\centering
\fontsize{8}{10}\selectfont
\setlength{\tabcolsep}{3pt}
\caption{Performance Contribution of Language-Agnostic vs. Language-Specific Semantic Analyzer.}
\label{tab:universal-robustness}
\resizebox{\columnwidth}{!}{%
\begin{tabular}{l cc cc}
\toprule
\multirow{2}{*}{\textbf{Language}} & \multicolumn{2}{c}{\textbf{Soundness}} & \multicolumn{2}{c}{\textbf{Completeness}} \\
\cmidrule(lr){2-3} \cmidrule(lr){4-5}
& \textbf{Agnostic-only} & \textbf{YASA} & \textbf{Agnostic-only} & \textbf{YASA} \\
\midrule
Java & 78 (\textcolor{red}{\textbf{2\%$\downarrow$}}) & 80 (72\%) & 31 (\textcolor{red}{\textbf{2\%$\downarrow$}}) & 32 (55\%) \\
Go & 80 (\textcolor{red}{\textbf{15\%$\downarrow$}}) & 96 (91\%) & 41 (\textcolor{red}{\textbf{6\%$\downarrow$}}) & 45 (66\%) \\
JavaScript & 107 (\textcolor{red}{\textbf{8\%$\downarrow$}}) & 118 (88\%) & 33 (\textcolor{red}{\textbf{4\%$\downarrow$}}) & 35 (71\%) \\
Python & 145 (\textcolor{red}{\textbf{12\%$\downarrow$}}) & 176 (70\%) & 32 (\textcolor{red}{\textbf{16\%$\downarrow$}}) & 44 (59\%) \\
\bottomrule
\end{tabular}%
}
\end{table}

\noindent\textbf{Engineering Effort to Support Each Language.} We evaluate the reduction in engineering complexity achieved by YASA's language-agnostic semantic analyzer by quantifying how much analysis logic can be shared across different programming languages. As shown in Table~\autoref{tab:semantic-reuse}, the language-agnostic semantic analyzer handles an average of 77.3\% of all semantic operations through 52 shared processing functions. These functions implement key analysis tasks such as variable resolution, expression evaluation, control flow analysis, and memory state management. This design ensures that language-specific analyzers only need to implement unique semantic characteristics rather than reimplementing core analysis algorithms.  
For instance, JavaScript requires 19 additional specialized semantic handlers, Python 15, Java 10, and Go 18, representing an incremental effort of only 16.1\% to 26.8\% across languages. We demonstrate that these shared semantics significantly reduce the engineering effort required to add support for new languages, validating the extensibility of its unified design.

\noindent\textbf{Robustness of Language-Agnostic Analyzer.} We assess the robustness of the language-agnostic semantic analyzer by evaluating its standalone performance across different programming languages without language-specific enhancements. This evaluation reveals how well language-agnostic semantic rules generalize across diverse programming paradigms. As shown in \autoref{tab:universal-robustness}, when we evaluate using only the language-agnostic semantic analyzer (effectively disabling language-specific handlers), YASA still correctly solves a significant portion of test cases. It achieves a 77\% success rate for JavaScript (140/183), 70\% for Go (121/173), 65\% for Java (109/169), and 54\% for Python (177/326). This performance demonstrates that the language-agnostic semantic rules capture essential analysis patterns across languages. The results also highlight the crucial role of language-specific handlers, which increase overall benchmark success from 64\% (547/851) to 74\% (626/851). 

\subsection{RQ3: Real-world Practicality}

To evaluate YASA's practical effectiveness in real-world deployment scenarios, we conducted a large-scale study at Ant Group, focusing on both vulnerability discovery capabilities and scanning efficiency.

\begin{table}[t]
\centering
\fontsize{8}{10}\selectfont
\setlength{\tabcolsep}{3pt}
\caption{Vulnerability Discovery at Ant Group.}
\label{tab:vulnerability-discovery}
\resizebox{\columnwidth}{!}{%
\begin{tabular}{llcccc}
\toprule
\textbf{Language} & \textbf{Type} & \textbf{Reported} & \textbf{Sanitized} & \textbf{Conf.} & \textbf{Fixed} \\
\midrule
\multirow{4}{*}{\textbf{Python}} 
& CMDi & 29 & 18 & 6 & 4 \\
& CODEi & 57 & 30 & 13 & 9 \\
& SSRF & 93 & 46 & 31 & 24 \\
& Deserialization & 3 & / & 3 & 2 \\
\midrule
\multirow{3}{*}{\textbf{Go}} 
& CMDi & 7 & 5 & 2 & 2 \\
& SQLi & 28 & 19 & 9 & 9 \\
& SSRF & 29 & 17 & 11 & 10 \\
\midrule
\multirow{2}{*}{\textbf{JavaScript}} 
& HPE & 55 & 36 & 12 & 11 \\
& SSRF & 13 & 8 & 5 & 5 \\
\midrule
\textbf{Total} & \textbf{/} & \textbf{314} & \textbf{179} & \textbf{92} & \textbf{76} \\
\bottomrule
\end{tabular}%
}
\end{table}

\noindent\textbf{Vulnerability Discovery.} We deployed YASA in production environments at Ant Group, scanning approximately 7.3K applications with over 100 million lines of multi-language code. The target systems span various web services, including frameworks like Egg.js, Flask, Spring, and Gin. As shown in \autoref{tab:vulnerability-discovery}, YASA identified 314 previously unknown vulnerability paths across 6 different vulnerability categories, including command injection, SQL injection, server-side request forgery (SSRF), privilege escalation, and deserialization attacks. Security experts at Ant Group confirmed 92 of them (29.3\%) as 0-day vulnerabilities, and 76 have already been fixed by the developers. 
As for the false positives, our manual verification reveals that 80.6\% of unconfirmed taint paths (179 out of 222 false positives) were attributed to the sanitization mechanisms along the taint paths that YASA’s current implementation does not fully recognize.

\noindent\textbf{Performance Efficiency.} 
To evaluate YASA's efficiency, we randomly sampled 20 applications for each supported language from the production environment. \autoref{tab:performance} details the performance metrics. YASA achieved an average scanning speed of 31.8 KLOC/min across the sampled dataset. For comparison, on the same dataset, industry-standard frameworks CodeQL and Joern achieved throughputs of 9.3 KLOC/min and 17.1 KLOC/min, respectively. Notably, Java analysis demonstrated the highest efficiency at 52.1 KLOC/min, benefiting from the robust type system and UAST optimization. Even for dynamic languages like Python (27.4 KLOC/min) and JavaScript (9.2 KLOC/min), the analysis time remains within acceptable limits for daily development workflows. These results confirm that YASA balances deep semantic analysis with the high-speed requirements in large-scale industrial deployment.

\begin{table}[t]
\centering
\fontsize{8}{10}\selectfont
\setlength{\tabcolsep}{4pt}
\caption{Performance Evaluation of YASA on 80 Randomly Sampled Production Applications at Ant Group.}
\label{tab:performance}
\resizebox{\columnwidth}{!}{%
\begin{tabular}{lrrrrr}
\toprule
\textbf{Language} & \textbf{\#Proj.} & \textbf{\#LoC} & \textbf{Time (s)} & \textbf{Avg Time (s)} & \textbf{\#KLoC/min} \\
\midrule
Java & 20 & 3,945,483 & 4,548.1 & 227.4 & 52.1 \\
Python & 20 & 2,187,668 & 4,782.7 & 239.1 & 27.4 \\
Go & 20 & 1,432,519 & 2,372.2 & 118.6 & 36.2 \\
JavaScript & 20 & 561,946 & 3,650.2 & 182.5 & 9.2 \\
\midrule
\textbf{Average} & \textbf{/} & \textbf{2,031,904} & \textbf{3,838.3} & \textbf{191.9} & \textbf{31.8} \\
\bottomrule
\end{tabular}%
}
\end{table}
\section{Discussion}

\textbf{Limitations.} While YASA demonstrates potential for multi-language taint analysis, several limitations remain. 
First, the UAST specification is derived from the characteristics of the currently supported languages. Supporting future languages with substantially different paradigms may require extending the UAST to capture new semantic constructs.
Second, the evaluation in RQ1 relies on xAST, a synthetic micro-benchmark, rather than macro-benchmarks with real-world vulnerabilities. This may limit external validity. We adopted this design to reduce biases from benchmark-specific vulnerability patterns and taint specifications.
Third, YASA is not fully sound framework. Instead, it follows a soundiness-oriented design~\cite{livshits2015soundiness} to balance precision and performance in practice. In particular, loops are typically handled through bounded unrolling rather than fixpoint computation, which may miss vulnerabilities involving complex control flow, such as deeply nested loops or recursion.
Finally, YASA models common built-in functions and widely used third-party libraries, but applies conservative taint propagation to unknown functions. While this design improves robustness, it can introduce false positives when unknown functions actually sanitize or validate tainted data.

\noindent\textbf{Future Work.} Several promising research directions could significantly expand its impact and capabilities.
First, YASA's extensible architecture provides a natural foundation for supporting emerging programming languages~(such as Rust, Ruby, ArkTS, etc.) and modern frameworks~(React, Django, Vue, etc.) that exhibit unique computational semantics. 
Second, YASA's unified multi-language representation opens opportunities for cross-language analysis that maintains semantic precision across language boundaries. Unlike traditional approaches that perform separate single-language analyses and subsequently bridge results through interface matching, YASA's unified execution model can preserve calling context sensitivity and data flow relationships across multi-language applications. 
Third, YASA could benefit from incorporating classical techniques such as fixpoint computation and sparse value-flow analysis to improve its universal semantic computation processes, addressing both soundness limitations and computational efficiency challenges.
Finally, we plan to develop a unified query language (QL) interface for YASA, which could enable declarative vulnerability rule specification and reduce migration barriers for organizations currently using CodeQL-based security workflows.
\section{Related Works}

\noindent \textbf{Single-Language Static Taint Analysis.}
Static taint analysis has been extensively studied for individual programming languages, leading to many language-specific tools. For Java, mature systems such as FlowDroid~\cite{arzt2014flowdroid}, Amandroid~\cite{wei2014amandroid}, DroidSafe~\cite{gordon2015information}, TAJ~\cite{tripp2009taj}, and Tai-e~\cite{tan2023taie} provide strong support for pointer and taint analysis in mobile and web applications. For C/C++, tools such as SVF~\cite{sui2016svf,cheng2024csa} and Pinpoint~\cite{shi2018pinpoint} address scalable memory and value-flow analysis. For JavaScript, prior work includes TAJS~\cite{simon2009tajs,laursen2024approximate}, DoubleX~\cite{fass2021doublex}, ODGen~\cite{li2022odgen,li2021proto,kang2023fast}, and GraphJS~\cite{ferreira2024graphjs}, which tackle challenges such as dynamic typing, prototype-based dispatch, and object dependency tracking. For Python, tools such as PySA~\cite{pysa} and PyT~\cite{pyt} rely on type inference to support taint analysis under dynamic typing.
Recent studies have also explored LLM-assisted approaches, such as LLMDFA~\cite{wang2024llmdfa}, RepoAudit~\cite{guo2024repoaudit}, and Artemis~\cite{ji2025artemis}. Despite their effectiveness, these approaches remain largely language-specific and therefore require substantial engineering effort to support multi-language analysis.

\noindent \textbf{Multi-Language Taint Analysis.}
Several frameworks have been proposed for multi-language static analysis. CodeQL~\cite{codeql,moor2007ql} supports cross-language querying through language-specific extractors and databases, but adding new languages requires substantial engineering effort and may introduce semantic inconsistencies. Joern~\cite{fabian2014joern,michael2017phpjoern} maps multiple languages to unified CPGs, but preserving language-specific semantics remains challenging. WALA~\cite{wala,santos2022wala} uses a common intermediate representation and mainly targets JVM-based languages, limiting its applicability to languages with different execution models. Infer~\cite{infer} shares analysis infrastructure across frontends but focuses primarily on memory safety for compiled languages, while Semgrep~\cite{semgrep} emphasizes parsing-based analysis with limited support for inter-procedure analysis. Overall, existing frameworks still struggle to balance unified representations with precise modeling of diverse language semantics.

\section{Conclusion}
In this paper, we presented YASA, a unified multi-language static taint analyzer. YASA introduces the UAST representation combined with point-to analysis to achieve high-precision taint analysis across multiple programming languages. Our comprehensive evaluation demonstrates that YASA consistently outperforms existing single-language and multi-language frameworks. The real-world deployment across 7,300 multi-language applications resulted in the discovery of 92 confirmed 0-day vulnerabilities, demonstrating YASA's practical impact for industrial software security. 

\section*{Acknowledgment}
This work was supported in part by the National Natural Science Foundation of China (grants No.62572209, 62502168), the Hubei Provincial Key Research and Development Program (grant No. 2025BAB057), and Ant Group Research Fund.

\balance
\bibliographystyle{ACM-Reference-Format}
\bibliography{ref}

\end{document}